\documentclass[prl,twoside,amsfonts,amssymb,amsmath,floatfix
,letterpaper,twocolumn,showpacs]{revtex4}
\usepackage{amsmath}
\usepackage{amsfonts}
\usepackage{amssymb}
\usepackage{graphicx}
\usepackage{graphicx,array}
\usepackage{dcolumn}
\usepackage{bm}

\begin{document}
\title{Separating the classical and quantum information via quantum cloning }
\author{M. Ricci$^{1}$, F. Sciarrino$^{1},$ N.J. Cerf$^{2}$, R. Filip$^{3}$, J.
Fiur\'{a}\v{s}ek$^{3}$, and F. De Martini$^{1}.$}
\affiliation{$^{1}$Dipartimento di Fisica and Istituto Nazionale per la Fisica della
Materia, Universit\`{a} di Roma ''La Sapienza'', p.le A. Moro 5, Roma, I-00185, Italy\\
$^{2}$Ecole Polytechnique, Universit\.{e} Libre de Bruxelles, Bruxelles,
B-1050, Belgium \\
$^{3}$ Department of Optics, Palack\'{y} University, 17. listopadu
50,Olomouc 77200, Czech Republic }

\date{\today}

\begin{abstract}
An application of quantum cloning to optimally interface a quantum system
with a classical observer is presented, in particular we describe a
procedure to perform a minimal disturbance measurement on a single qubit by
adopting a $1\rightarrow 2$ cloning machine followed by a generalized
measurement on a single clone and the anti-clone or on the two clones. Such
scheme has been applied to enhance the transmission fidelity over a lossy
quantum channel.
\end{abstract}

\maketitle
Information is a property of physical systems, that can be defined and
quantified within any physical model. While basic principles are assumed to
be generally valid, a coherent analytic formulation of an information theory
is deeply related to the modalities by which the knowledge about a system is
acquired by a (classical) observer. In quantum theory an observer cannot
extract all the information about an unknown state $\left| \phi
\right\rangle $ by a measurement performed on a finite ensemble of
identically prepared systems. In particular, the mean fidelity $G$ of any
state estimation strategy based on the measurement of N copies of a qubit $%
\left| \phi \right\rangle $ must satisfy the bound $G\leq
G_{opt}=(N+1)/(N+2) $ \cite{Mass95}$,$ where $G$ is defined as the\ mean
overlap between the unknown state $\left| \phi \right\rangle $ and the state
inferred from the measurement $\rho _{G}$: $G=\left\langle \phi \right| \rho
_{G}\left| \phi \right\rangle $. Moreover, any gain of knowledge
irreversibly alters the estimated system. Recently the disturbance
associated to the estimation process has been characterized analytically for
a generic $d$-level quantum system and the optimal ratio between the
classical information acquired, $G,$ and the quantum fidelity $%
F=\left\langle \phi \right| \rho _{S}\left| \phi \right\rangle $ of the
output state $\rho _{S}$ has been found by Banaszek \cite{Bana01} (Fig.1-a).
\begin{figure}
\includegraphics[height=8cm,angle=-90]{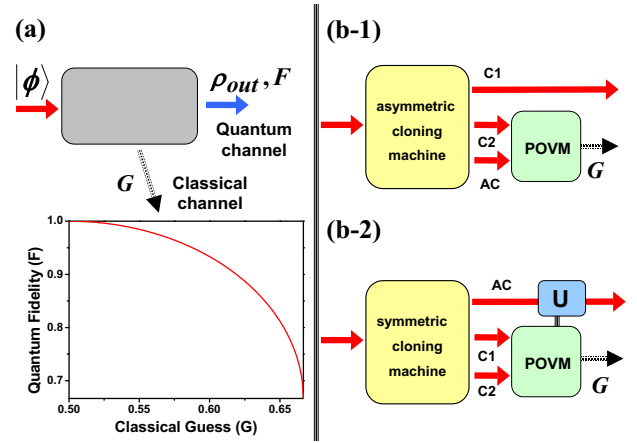}
\caption{({\bf a}) Plot of the optimal quantum fidelity versus the
classical guess of the state; ({\bf b}) Schematic diagrams of a minimal
disturbance measurement on a single qubit performed by adopting: ({\bf 1}%
)-an asymmetric cloning machine and a POVM; ({\bf 2})- a symmetric cloning
machine, a POVM and a classical feed-forward.} \label{fig:1}
\end{figure}
These fundamental results of the classical-quantum interface theory also
affect the quantum process of the distribution of information from a single
quantum system to many ones: one of the obvious consequences of the bound on
the fidelity of estimation is that unknown states of quantum systems cannot
be perfectly copied \cite{Woot82}. Certainly if this would be possible, then
one would be able to violate the boundary value $G_{opt}$. The problem of
manipulating and controlling the flux of quantum information has been in
general tackled and solved by the theory of quantum cloning \cite{Buze96}.
Actually the optimal cloning processes generate copies which exhibit the
maximum values of quantum fidelity achievable in compliance to quantum
mechanics rules; this feature renders such devices an essential instrument
for the assessment of the security of quantum cryptographic protocols \cite
{Cerf02}.

In this work we show that quantum cloning is a fundamental tool not only for
the distribution of quantum information but also to interface a quantum
system with a classical observer, that is, to optimally split the original
information content associated with any system into a classical and a
quantum contributions. Precisely, a minimal disturbance measurement on a
qubit can be implemented adopting a $1\rightarrow 2$ universal cloning
machine followed by a proper POVM on the two clones or on a single clone and
the anti-clone. The minimal disturbance implies a measurement that saturates
the quantum mechanical trade-off between the information gained by the
observer and the quantum state disturbance induced by the estimating process.%
{\large \ }Two different strategies will be illustrated: the first one
exploits a tunable asymmetric cloning machine followed by a fixed POVM while
the second one employs a symmetric cloning machine, a variable POVM and a
classical feed-forward in analogy with the teleportation protocol \cite
{Benn93}.

Let us consider a single ($N=1$) qubit, e.g. encoded in the polarization of
a single photon. The expression of the Banaszek's bound reads: 
\begin{equation}
\sqrt{F-\frac{1}{3}}\leq \sqrt{G-\frac{1}{3}}+\sqrt{\frac{2}{3}-G}
\label{BanaBound}
\end{equation}

For the sake of simplicity, we restrict our considerations to the $%
N=1\rightarrow M=2$ asymmetric optimal quantum cloning machine (AQCM) \cite
{Cerf00,Fili04} which generates two clones $C1$ and $C2$ with different
fidelities $F_{C1}$ and $F_{C2}$\ starting from the input qubit in the state 
$\left| \phi \right\rangle $ and from two ancillas. For an asymmetric
universal cloner, a simple way to express the transformation acting on $%
\left| \phi \right\rangle $ is 
\begin{equation}
\left| \phi \right\rangle \rightarrow \nu \left| \phi \right\rangle
_{C1}\left| \Psi ^{-}\right\rangle _{C2,AC}+\mu \left| \phi \right\rangle
_{C2}\left| \Psi ^{-}\right\rangle _{C1,AC}  \label{AQCM}
\end{equation}
where $AC$ denotes the third qubit, usually called ''anticlone'' and $\left|
\Psi ^{-}\right\rangle =2^{-1/2}(\left| 01\right\rangle -\left|
10\right\rangle )$. Here $\left| \mu \right| ^{2}$ is the depolarizing
fraction of clone $C1$, that is, the probability that $C1$ is depolarized,
so that the corresponding fidelity reads $F_{C1}(\mu )=1-\left| \mu \right|
^{2}/2.$ Of course, $\left| \nu \right| ^{2}$ is the depolarizing fraction
of clone $C2$, and we have similar expressions $F_{C2}(\nu )=1-\left| \nu
\right| ^{2}/2$. The normalization condition reads $\left| \mu \right|
^{2}+\mu \nu +\left| \nu \right| ^{2}=1.$ Note that ($\mu =0;\nu =1$) is a
trivial cloner where the original is transferred to $C1$, while $C2$ is
random. The value $\mu =\nu =1/\sqrt{3}$ corresponds to the symmetric
machine, hence $F_{C1}=F_{C2}=5/6.$ The cloning transformation (\ref{AQCM})
can be re-expressed in the following form 
\begin{widetext}
\begin{equation}
\left| \phi \right\rangle \rightarrow \left( \frac{\mu }{2}+\nu \right)
\left\{ \left| \phi \right\rangle _{C1}\left| \Psi ^{-}\right\rangle
_{C2,AC}\right\} +\frac{\mu }{2}\left\{ \sigma _{Z}\left| \phi \right\rangle _{C1}\left|
\Psi ^{+}\right\rangle _{C2,AC}+\sigma _{X}\left| \phi \right\rangle
_{C1}\left| \Phi ^{-}\right\rangle _{C2,AC}+\sigma _{Y}\left| \phi
\right\rangle _{C1}\left| \Phi ^{+}\right\rangle _{C2,AC}\right\}
\label{AQCM2}
\end{equation}
\end{widetext}
where $\left| \Psi ^{\pm }\right\rangle =2^{-1/2}(\left| 01\right\rangle \pm
\left| 10\right\rangle )$ and $\left| \Phi ^{\pm }\right\rangle
=2^{-1/2}(\left| 00\right\rangle \pm \left| 11\right\rangle )$ are the four
Bell states of the qubits $C2$ and $AC.$ A straightforward conclusion we can
draw from this expression is that performing a Bell measurement on $C2$ and $%
AC$ gives all the information needed to perfectly reconstruct the original
state $\left| \phi \right\rangle $ from $C1$ \cite{Brus01}. Just like in
teleportation, we need to apply one of the Pauli operators on $C1$ depending
on the outcome of the Bell measurement. Also, by tracing over $C2$ and $AC$,
we see that the clone $C1$ is left in the state $\rho _{C1}=(1-\left| \mu
\right| ^{2})\left| \phi \right\rangle \left\langle \phi \right| +\left| \mu
\right| ^{2}{\mathbb I}/2.$ Similar conclusions can be obtained for the clone $%
C2$ and the anticlone $AC$, which is in state $\mu \nu |\phi _{\perp
}\rangle \langle \phi _{\perp }|+(|\mu |^{2}+|\nu |^{2}){\mathbb I}/2$, where $%
|\phi _{\perp }\rangle $ denotes a state orthogonal to $|\phi \rangle $, $%
\langle \phi _{\perp }|\phi \rangle =0$.

{\bf Asymmetric cloning - }The basic idea of this work is the following: the
quantum information carried by the input system $\left| \phi \right\rangle $
is distributed into a larger number of qubits adopting the AQCM and then,
while the clone $C1$ contains an approximate replica of $\left| \phi
\right\rangle $ quantified by the fidelity $F_{C1}=F$, the other outputs of
the machine, $C2$ and $AC$, are coherently measured to acquire classical
information on the initial state and estimate it with fidelity $G$
(Fig.1-b1).The information preserving property of the cloning process
suggests that the optimal value of $G$ achievable by this procedure should
satisfy the Banaszek bound for the given value of $F$. Intuitively the
optimal procedure could consist of the state estimation of a state $\left|
\phi \right\rangle $ from a pair of orthogonal qubits $|\phi \rangle |\phi
_{\perp }\rangle $.

Let us first establish notation for the POVM. The optimal covariant POVM for
the estimation of $|\phi \rangle $ from a single copy of $|\phi \rangle
|\phi _{\perp }\rangle $ has the structure 
\begin{equation}
\Pi (\Omega )=U(\Omega )\otimes U(\Omega )\Pi _{0}U^{\dagger }(\Omega
)\otimes U^{\dagger }(\Omega ),  \label{POVMorthogonal}
\end{equation}
where the unitary $U(\Omega )$ generates the states $|\Omega \rangle $ and $%
|\Omega _{\perp }\rangle $ from the computational basis states, $|\Omega
\rangle =U(\Omega )|0\rangle $ and $|\Omega _{\perp }\rangle =U(\Omega
)|1\rangle .$ The POVM $\Pi (\Omega )$ must satisfy the normalization
condition, $\int_{\Omega }\Pi (\Omega )d\Omega ={\mathbb I},$where ${\mathbb I}$
denotes the identity operator and $d\Omega $ is the invariant Haar measure
on the group SU(2). If the measurement result is $\Pi (\Omega )$, then the
estimated state reads $U(\Omega )|0\rangle $. The operator $\Pi _{0}$ which
generates the optimal POVM (\ref{POVMorthogonal}) has rank one and can be
expressed as $\Pi _{0}=|\pi _{0}\rangle \langle \pi _{0}|$, where $|\pi
_{0}\rangle =\frac{\sqrt{3}+1}{\sqrt{2}}\left[ |01\rangle +(2-\sqrt{3}%
)|10\rangle \right] $. The covariant POVM is continuous but it can be
discretized by choosing only several particular $\Omega _{j}\equiv
(\vartheta _{j},\phi _{j})$ such that $\sum_{j}\Pi (\Omega _{j})\propto 
{\mathbb I}$. As found by \cite{Gisi99} an optimal strategy consists in the
following discrete POVM $\left| \Theta _{i}\right\rangle \left\langle \Theta
_{i}\right| ,\{i=1,4\} $
\[
\left| \Theta _{i}\right\rangle =\gamma \left| \overrightarrow{n_{i}},-%
\overrightarrow{n_{i}}\right\rangle -\delta \sum_{k\neq i}\left| 
\overrightarrow{n_{k}},\overrightarrow{-n_{k}}\right\rangle 
\]
where $\gamma =13/(6\sqrt{6}-2\sqrt{2})$, $\delta =(5-2\sqrt{3})/(6\sqrt{6}-2%
\sqrt{2}),$ $\left\{ \overrightarrow{n}_{i}\right\} $ represents the
directions of the four vertices of a tethraedron in the Bloch sphere with
the following cartesian coordinates $\overrightarrow{n_{1}}=\left(
0,0,1\right) ,$ $\overrightarrow{n_{2}}=\frac{1}{3}\left( \sqrt{8},0,-1\right) ,$ $\overrightarrow{n_{3}}=\frac{1}{3}\left( -\sqrt{2},\sqrt{6},-1\right) ,$ $\overrightarrow{n_{4}}=\frac{1}{3}\left( -\sqrt{2},-\sqrt{6},-1\right) .$

The initial state of the quantum system $C2$-$AC$ in the basis $\left\{
\left| \phi \right\rangle ,\left| \phi ^{\bot }\right\rangle \right\} $ is
expressed by the density matrix $\rho _{C2,AC},$ attained by tracing over
the system $C1$ in Eq.\ref{AQCM}. Applying the POVM $\left| \Theta
_{i}\right\rangle \left\langle \Theta _{i}\right| $ the output $i$ is
obtained with probability $p_{i}=tr(\left| \Theta _{i}\right\rangle
\left\langle \Theta _{i}\right| \rho _{C2,AC})$ and the input qubit is
guessed to be in the state\ $\left| \overrightarrow{n_{i}}\right\rangle .$
The amount of classical information about $\left| \phi \right\rangle $
attained is $G(\left| \phi \right\rangle )=\sum_{i}p_{i}\left| \left\langle
\phi \right. \left| \overrightarrow{n_{i}}\right\rangle \right| ^{2}$ and
the average value of $G$ over all possible input states is equal to $%
G=\int_{\left| \phi \right\rangle \epsilon H}G(\left| \phi \right\rangle
)d\phi .$ From the previous expressions we obtain for $F$ and $G$ the
functional relation 
\begin{equation}
F(G)=\frac{1}{3}+\left( \sqrt{G-1/3}+\sqrt{-G+2/3}\right) ^{2}
\label{banabound}
\end{equation}
that saturates the Banaszek's bound \cite{Bana01}. Note that the optimal
measurement on the second clone and anticlone does not depend on the
asymmetry of the cloner. The latter is only used here to tune the balance
between $F$ and $G$.

{\bf Symmetric cloning - }Let us now investigate whether the optimal
trade-off between $F$ and $G$ can be obtained by varying the measurement on
the two clones generated through the symmetric cloning machine. In this case
the optimal covariant POVM is generated by the rank-one operator $\widetilde{%
\Pi }_{0}=|\widetilde{\pi }_{0}\rangle \langle \widetilde{\pi }_{0}|$, where 
$|\widetilde{\pi }_{0}\rangle =\xi |00\rangle +\sqrt{3-\xi ^{2}}|11\rangle .$
This measurement interpolates between optimal POVM for state estimation from 
$|\phi \rangle |\phi \rangle $ ($\xi =\sqrt{3}$) leading to the maximum
value $G_{opt}=\frac{2}{3}$ and the Bell measurement in the basis of
maximally entangled states ($\xi =\sqrt{3/2}$) leading to $F=1$ through a
reversion strategy.

Let us calculate the estimation fidelity for the covariant POVM generated by 
$\widetilde{\Pi }_{0}$. The mean fidelity can be calculated by averaging
over all input states and over the POVM. However, since the POVM is
covariant, it suffices to consider only a single input state, e.g.,$%
|0\rangle $. By exploiting the expression (\ref{AQCM}) for $\mu =1/\sqrt{3}$
and $|\phi \rangle =|0\rangle $we obtain the average fidelity $%
G=\int_{\Omega }tr(\widetilde{\Pi }(\Omega )\rho _{C1,C2})\left|
\left\langle 0\right| U(\Omega )|0\rangle \right| ^{2}d\Omega $ where $\rho
_{C1,C2}$ is the reduced density matrix of systems $C1-C2.$ The final result
is $G=\frac{1}{3}+\frac{\xi ^{2}}{9}.$\ If the measurement result is $%
\widetilde{\Pi }(\Omega )$, then the correcting unitary $U(\Omega
)U^{T}(\Omega )\sigma _{Y}$ should be applied to the anti-clone $C$. The
mean fidelity between the anti-clone after this correction and the input
state $|\phi \rangle $ can be evaluated as $F=\frac{2}{3}+\frac{2}{9}\xi 
\sqrt{3-\xi ^{2}}.$ If we express $\xi $ in terms of $G$, we find that $F(G)$
is equal to expression (\ref{banabound}). This proves that the Banaszek's
bound is saturated. We have assumed here that the optimal POVM is covariant
and continuous but we could of course discretize it and find an equivalent
POVM with finite number of elements.

{\bf Applications -\ }In the present paragraphs, we shall exploit the
quantum cloning to improve a simple quantum communication task. Let us
consider the following problem: Alice wants to transmit an unknown quantum
state $\left| \phi \right\rangle $ encoded into a single photon to Bob
through a lossy channel (Fig.2-a). The quantum communication channel is
characterized by the transmittivity $p$, i.e. the probability that the
photon reaches Bob's station. In the case in which Alice directly sends the
photon to Bob, the fidelity of the quantum state transmission is found to be 
$F_{dir}=(1+p)/2$ (Fig.2-b). Indeed when the qubit reaches Bob, event which
occurs with probability $p$, the fidelity of transmission is equal to $1$,
otherwise, when the qubit is lost, Bob must guess randomly the quantum state
of $\left| \phi \right\rangle $ and the fidelity is equal to $\frac{1}{2}$.
In order to enhance this transmission fidelity we shall investigate
different alternative strategies based on the cloning process.
\begin{figure}
\includegraphics[width=6.5cm,height=8cm]{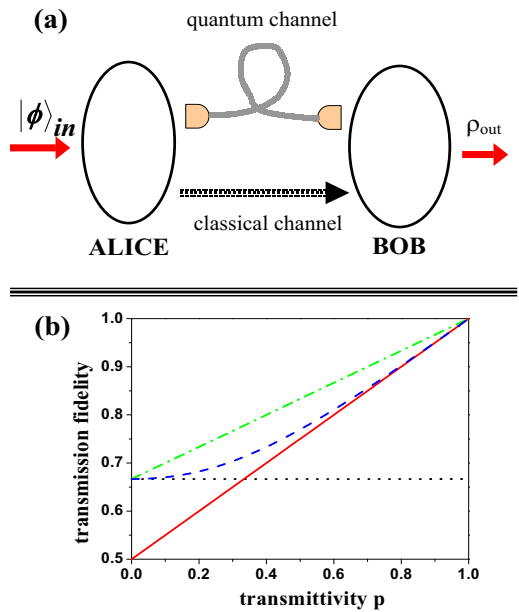}
\caption{ ({\bf a}) Schematic diagram of a generic communication channel; (%
{\bf b})\ Fidelity of the transmission of a quantum state through a quantum
channel characterized by a transmittivity $p$: solid line ($F_{dir}$),
dashed line ($F_{cl}$), dotted line ($G_{opt}$) and dashed-dotted line ($%
F_{QM}$)} \label{fig:2}
\end{figure}
In a first scenario involving the asymmetric cloner, the clone $C1$ is sent
down the quantum channel while the qubits $C2$ and $AC$ are kept at the
sender station (Fig.1-b1). Alice optimally estimates the input state with
fidelity $G$ by performing the POVM\ $\Pi (\Omega )$\ on $C2$ and $AC$ (Eq.(%
\ref{POVMorthogonal})) and communicates the result to Bob. Let us first note
that if no memory is available and the measurement on sender's side is
independent of whether the state was delivered to receiver or lost in the
channel, then the Banaszek bound applies and cannot be beaten. The overall
transmission fidelity is now $F_{C1}$ when the qubit reaches Bob and $G$
when Bob is forced to exploits the classical information, since the photon
is lost. Hence the average fidelity reads $F(p)$=$pF_{C1}+(1-p)G$. By
optimizing the asymmetry of the cloning machine, that is the parameter $\mu $%
, with respect to the transmittivity $p$ of the channel we obtain (Fig.2-b) $%
F_{cl}(p)=\frac{1}{6}\left( 3+p+\sqrt{1+p(5p-2)}\right) .$ This is the
optimal strategy based on a classical-quantum communication since the
present procedure saturates the Banaszek's bound, as said. Similar result
can be obtained by adopting a symmetric cloning machine at the sender
station. In this case the $POVM$ $\widetilde{\Pi }(\Omega )$ is performed on
the two clones (Fig.1-b2): depending on the result Alice applies the
appropriate feed-forward $U$ to the AC qubit and then sends it to Bob.

An higher fidelity of transmission can be obtained by a more sophisticated
approach. Let us suppose that Alice can use a quantum memory \cite{Juls04}
whereas Bob can communicate to her whether or not he has received the
transmitted photon. If the photon reaches Bob's site, they apply a reversion
procedure and recover the initial qubit $\left| \phi \right\rangle $ at
Bob's station. We can apply two different strategies which leads to the same
fidelity of transmission (Fig.2-b) $F_{qm}=\frac{2}{3}+\frac{1}{3}p$. In the
first approach Alice employs a symmetric cloning and transmits the anticlone
to Bob. If the photon is lost, Alice performs an optimal estimation on the
clones achieving a fidelity $\frac{2}{3},$ otherwise she carries out an
incomplete\ Bell measurement on the two clones and sends her results to Bob
that applies the appropriate unitary Pauli operator to the qubit $AC$ in
order to recover $\left| \phi \right\rangle $. Since the two clones belong
to the symmetric subspace 1 trit of information must be transmitted from
Alice to Bob. The quantum memory is necessary since Alice must wait Bob's
message to decide whether she implements a Bell measurement or an estimation
POVM. We note that the same transmission fidelity can be achieved by
adopting the standard teleportation protocol over a lossy quantum channel
with transmission probability $p$.

The quantum cloning can be also used to protect from losses and decoherence
a state stored in a quantum memory. Consider a simple model where a qubit
stored in a memory is preserved with probability $p$ and is erased with
probability $1-p$ leading to a fidelity of storage $F_{S}=(1+p)/2$. Suppose
now that before storing we clone the state and keep in the memory both
clones as well as the anti-clone. If all three qubits are preserved, we
perfectly recover the state, otherwise if at least one clone is maintained
we get the fidelity $5/6$. If only the anti-clone is preserved, then we can
apply another approximate U-NOT gate and recover a state with fidelity $5/9$%
. Finally,when all qubits are lost we guess the state with fidelity $1/2$.
The average fidelity of this cloning-based strategy reads $%
F_{C}=(1+2p)(9-5p+2p^{2})/18$. Remarkably, $F_{C}-F_{S}$ is non-negative for
all $p\in \lbrack 0,1]$ so the cloning can partially protect the state in
the memory from the erasure. The improvement is maximum for $p=1/3$ when we
obtain $F_{C}-F_{S}=3.3\%$. 

{\bf Conclusions} - We have presented an explicit application of quantum
cloning to quantum-classical interface, in particular we described a
procedure to perform a minimal disturbance measurement. Such scheme can been
applied to enhance the transmission fidelity over a lossy (but noiseless)
channel. or to enhance the performance of a quantum memory with erasure.
These procedures exploit the cloning process to encode a single qubit into
the Hilbert space of three qubits. This redundancy, similar to that
exploited in quantum error correction codes, is the reason why the cloning
can help us in protecting quantum information from losses. We may expect
that the cloning can also help to protect against other kinds of
decoherence. However theoretical analysis reveals that the present strategy
doesn't work for Pauli channels. As a further application, we note that the
present method realizes an universal weak measurement \cite{Ahar88} in the
limit of vanishing disturbance. 

The implementation of the previous schemes adopting single photon states is
challenging but within the present technology. The cloning machine has been
realized either by an amplification process \cite{DeMa02,Lama02,Fili04} and
by linear optics techniques \cite{Ricc04,Fili04,Zhao04}, classical
feedforward has also recently been reported \cite{Giac02}, and the required
generalized measurements (POVMs) on two photonic qubits can be realized
probabilistically using linear optics

F.D.M., M.R., and F.S. acknowledge financial support from the FET European
Network IST-2000-29681 (ATESIT), the I.N.F.M. (PRA-CLON), and the Ministero
della Istruzione, dell'Universit\'{a} e della Ricerca (COFIN 2002). N.J.C.
acknowledges financial support from the Communaut\'{e} Fran\c{c}aise de
Belgique under grant ARC 00/05-251 and from the IUAP programme of the
Belgian government under grant V-18. R.F. and J.F. acknowledge support from
the EU under project SECOQC, from the grant MSM6198959213 of the Ministry of
Education of Czech Republic and from the project 202/03/D239 of the Grant
Agency of Czech Republic.

\end{document}